\documentclass[aps,prd,superscriptaddress,preprint,tightenlines,nofootinbib]{revtex4}

\usepackage{graphicx} 
\usepackage{dcolumn}  
\usepackage{bm}       

\def\1#1#2#3{\multicolumn{#1}{#2}{#3}}
\def \b{{\cal B}}
\def \bea{\begin{eqnarray}}
\def \beq{\begin{equation}}

\def \dof{{\rm dof}}
\def \eea{\end{eqnarray}}
\def \eeq{\end{equation}}

\def \minv{M_{\rm inv}}

\begin{document}

\preprint{CLNS 08/2037}       
\preprint{CLEO 08-20}         

\title{\boldmath Observation of $\chi_b(1P_J,2P_J)$ decays to light hadrons}




\author{D.~M.~Asner}
\author{K.~W.~Edwards}
\author{J.~Reed}
\affiliation{Carleton University, Ottawa, Ontario, Canada K1S 5B6}
\author{R.~A.~Briere}
\author{G.~Tatishvili}
\author{H.~Vogel}
\affiliation{Carnegie Mellon University, Pittsburgh, Pennsylvania 15213, USA}
\author{P.~U.~E.~Onyisi}
\author{J.~L.~Rosner}
\affiliation{Enrico Fermi Institute, University of
Chicago, Chicago, Illinois 60637, USA}
\author{J.~P.~Alexander}
\author{D.~G.~Cassel}
\author{J.~E.~Duboscq\footnote{Deceased}}
\author{R.~Ehrlich}
\author{L.~Fields}
\author{R.~S.~Galik}
\author{L.~Gibbons}
\author{R.~Gray}
\author{S.~W.~Gray}
\author{D.~L.~Hartill}
\author{B.~K.~Heltsley}
\author{D.~Hertz}
\author{J.~M.~Hunt}
\author{J.~Kandaswamy}
\author{D.~L.~Kreinick}
\author{V.~E.~Kuznetsov}
\author{J.~Ledoux}
\author{H.~Mahlke-Kr\"uger}
\author{D.~Mohapatra}
\author{J.~R.~Patterson}
\author{D.~Peterson}
\author{D.~Riley}
\author{A.~Ryd}
\author{A.~J.~Sadoff}
\author{X.~Shi}
\author{S.~Stroiney}
\author{W.~M.~Sun}
\author{T.~Wilksen}
\affiliation{Cornell University, Ithaca, New York 14853, USA}
% ...
\collaboration{CLEO Collaboration}
\noaffiliation

\date{August 7, 2008}

\begin{abstract} 
Analyzing $\Upsilon(nS)$ decays acquired with the CLEO detector operating at
the CESR $e^+e^-$ collider, we measure for the first time the product branching
fractions ${\cal B}[\Upsilon(nS)\to\gamma\chi_{b}((n-1)P_J)] \times
{\cal B}[\chi_{b}(n-1)P_J)\to X_i]$ for $n=2$ and $3$, where $X_i$
denotes, for each $i$, one of the fourteen exclusive light-hadron final states
for which we observe significant signals in both $\chi_b(1P_J)$ and
$\chi_b(2P_J)$ decays.  We also determine upper limits for the
electric dipole (E1) transitions $\Upsilon(3S) \to \gamma \chi_b(1P_J)$.
\end{abstract}

\maketitle

In the 31 years since the first observation of bottomonium we have learned a
great deal about decays of the $\Upsilon(1S,2S,3S)$ resonances and transitions
among them.  Less is known about $P$-wave states because they are not produced
directly in $e^+ e^-$ collisions. The spin-triplet $\chi_b$ mesons are produced
copiously in electric dipole (E1) transitions \cite{pdg}, permitting the
recent first observations of inclusive decays of $\chi_b(nP_J)$ to 
$p(\bar{p})$~\cite{barybb} and to open charm~\cite{opencharm}.
Nothing else is known about $\chi_b(nP_J)$ decays to non-$b\bar{b}$ states.
Such processes are of interest both intrinsically and as clues in searching for
states of mass $\sim 10$ GeV/$c^2$ via their exclusive decays.

In this article we report the first observations of decays
of $\chi_b(1P_J)$ and $\chi_b(2P_J)$ into specific final states of light
hadrons, where the $\chi_{b}(nP_J)$ states are produced via
$\Upsilon(2S) \to \gamma
\chi_b(1P_J)$ and $\Upsilon(3S) \to \gamma \chi_b(2P_J)$.
We also determine upper limits on
rates for the suppressed E1 transitions $\Upsilon(3S)\to\gamma \chi_b(1P_J)$.

We use the same
$\Upsilon(nS)$ on-resonance data as in
the analysis of Ref.~\cite{Artuso:2004fp}, corresponding to N$_{\Upsilon(nS)} =
(20.82 \pm0.37, 9.32\pm0.14, 5.88\pm0.10) \times 10^6$ resonance decays for
$n=1,~2$, and 3, respectively, collected by the CLEO III detector \cite{c3det}
at the the Cornell Electron Storage Ring.  Hadronic events are selected based
on the criteria used in the
analysis of Ref.~\cite{Artuso:2004fp}.

Our signal events have the form $\Upsilon(nS) \to \gamma X_i$, where $X_i$
denotes a specific fully reconstructed final state. We allow a large variety of
possibilities for $X_i$, but to keep the list finite and realistic we impose
the following requirements. Each $X_i$ consists of a combination of twelve or
fewer ``particles,'' where a ``particle'' is defined here to be a photon or a
charged pion ($\pi^\pm$), kaon ($K^\pm$), or proton ($p/\bar p$). Each state
$X_i$ must have at least two charged ``particles'' and conserve overall
charge, strangeness, and baryon number. We only consider modes in which photons
other than that from the transition are paired into either $\pi^0$ or $\eta$
candidates, of which we only permit four or fewer. Neutral kaon decays into
$\pi^+\pi^-$ are also considered.  With these criteria,
there are
659 separate final states, which act as the basis for our search.

Photon candidates are taken from calorimeter showers that do not match the
projected trajectory of any charged particle and which have a lateral shower
profile consistent with that of an isolated electromagnetic shower.  Each
candidate for a $\pi^\pm$, $K^\pm$, and $p/\bar p$ must be positively
identified as such by a combination of its specific ionization $dE/dx$, within
$3 \sigma$, where $\sigma$ refers to uncertainty
due to measurement errors, and, when available, the response of a Ring Imaging
Cherenkov system as in the analysis of Ref.~\cite{opencharm}.
Candidates for $\pi^0$ and $\eta$ decays to two photons are allowed only if the
photon pair mass is within $3 \sigma$ of the nominal $\pi^0$ or $\eta$ mass.
$K^0_S \to \pi^+ \pi^-$ candidates, consisting of a pair of vertex-constrained
oppositely charged tracks, are required to have effective mass within $3\sigma$
of the nominal mass~\cite{pdg} and to have a flight path
before decay exceeding twice the longitudinal vertex resolution.

We improve sample purity by constraining the transition photon plus the
decay products of $X_i$ to the initial $\Upsilon(nS)$ four-momentum with a 4C
kinematic fit and requiring the fitted $\chi^2(4C)/\dof < 5$
as in Ref.\ \cite{m1etac}.  The kinematic fit also allows us to improve the
resolution on the invariant mass of the $X_i$ by using fitted, instead of
measured, four-momenta: we denote this mass by $\minv$.

\begin{figure}
\begin{center}
\includegraphics[width=0.48\textwidth]{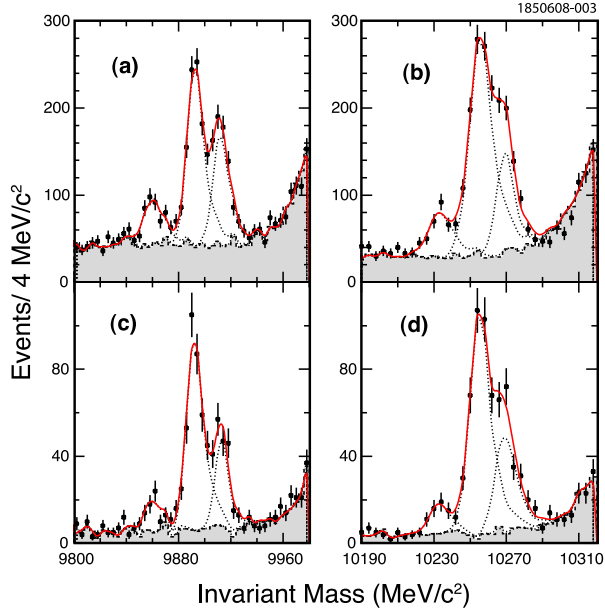}
\caption{$\minv$ spectra based on (a,c) $\Upsilon(2S)$ and (b,d) $\Upsilon(3S)$
data for the sum of all 659 modes (a,b) and the 14 selected modes (c,d).  The
observed peaks are consistent with the transitions 
$\Upsilon(2S)\to\gamma
\chi_{b}(1P_J)$ and $\Upsilon(3S)\to \gamma\chi_{b}(2P_J)$.
Fitted backgrounds are represented by dashed histograms, fitted $\chi_b(nP_J)$
peaks are represented by dotted lines, and sums of fitted signals and
background are denoted by solid curves.}
\label{fig:chibstacksum}
\end{center}
\end{figure}

Figure~\ref{fig:chibstacksum} shows fits to the $\minv$ distribution of the sum
of all 659 modes in (a) $\Upsilon(2S)$ and (b) $\Upsilon(3S)$ data.
The natural $\chi_b(nP_J)$ widths \cite{chibw} are expected to be much smaller
than the resolution ($\sim5$~MeV) of the transition photon, which has a mostly
Gaussian line shape and a low-energy tail induced by energy leakage out of the
crystals used in the algorithm.  This {\it Crystal Ball} line (CBL) shape is
discussed in more detail in Ref.\ \cite{CB}.  To fit the $\minv$ distribution
in Fig.\ \ref{fig:chibstacksum}, we use a ``reversed'' CBL shape with the
asymmetric tail on the high side instead of the low side of the peak.
The fitted
masses of $\chi_b(1P_J)$ and $\chi_b(2P_J)$ are consistent with the known
masses~\cite{pdg}.  For background shapes, we use $\minv$ spectra obtained with
the same analysis procedure but based on $\Upsilon(1S)$ data, {\it shifted}
by differences in center-of-mass energy while floating normalizations.
This procedure appears to represent backgrounds reasonably.
Using low-order polynomials instead of $\Upsilon(1S)$ data
to represent backgrounds, we obtain consistent results. 

With signal shapes, including central values, fixed by a fit to the sum of the
659 modes, we fit the unconstrained photon energy spectra for {\it each} mode
with CBL shapes.  We use unconstrained photon energy spectra because
calorimeter resolutions are independent of the final states in $\chi_b(1P,2P)$
decays.  We then determine significances from the fit to each mode as
$\sqrt{-2\ln (\mathcal{L}_{wo}/\mathcal{L}_{w})}$ where $\mathcal{L}_{wo}$ and
$\mathcal{L}_{w}$ are likelihoods from fits without and with an allowance for
signal.  We determine the significance from simultaneous fits to the three
peaks instead of determining the significance of individual 
$\chi_{b}(1P_J,2P_J)$
peaks.  We identify 14 modes giving at least $5 \sigma$ significance from
 {\it both}
$\chi_{b}(1P_J)$ and $\chi_{b}(2P_J)$ decays.

On the basis of \textsc{Geant}-based~\cite{GEANT} signal and various background
Monte Carlo (MC) samples for the 14 identified modes, the required limit on
$\chi^2(4C)/\dof$ is varied from its initial value of 5 in order to optimize
signal sensitivity while reducing backgrounds.  The optimum value for the 14
modes is found to be $\chi^2(4C)/\dof<3$, and is adopted as our nominal value.
As some modes show further improvement in sensitivity for $\chi^2(4C)/\dof<2$,
we also explore the choices $\chi^2(4C)/\dof<2$ and $<4$ in our 
study of systematic uncertainties.

Fig.\ \ref{fig:chibstacksum} shows $\minv$ distributions of (c) $\Upsilon(2S)$
and (d) $\Upsilon(3S)$ data based on the sum of the 14 modes with our nominal
selection criteria.  The fitted backgrounds are $\Upsilon(1S)$ data, shifted 
as in Figs.\ \ref{fig:chibstacksum}(a,b).  The fitted $\chi_b(nP_J)$ masses
again are consistent with the known values \cite{pdg}.
With the restriction of $\chi^2(4C)/\dof<3$, $\Upsilon(2S,3S)$ decays in the
14 modes lead to roughly 40$\%$ of the total observed events in the 659 modes. 
 
\begin{figure*}
\mbox{\includegraphics[width=0.48\textwidth]{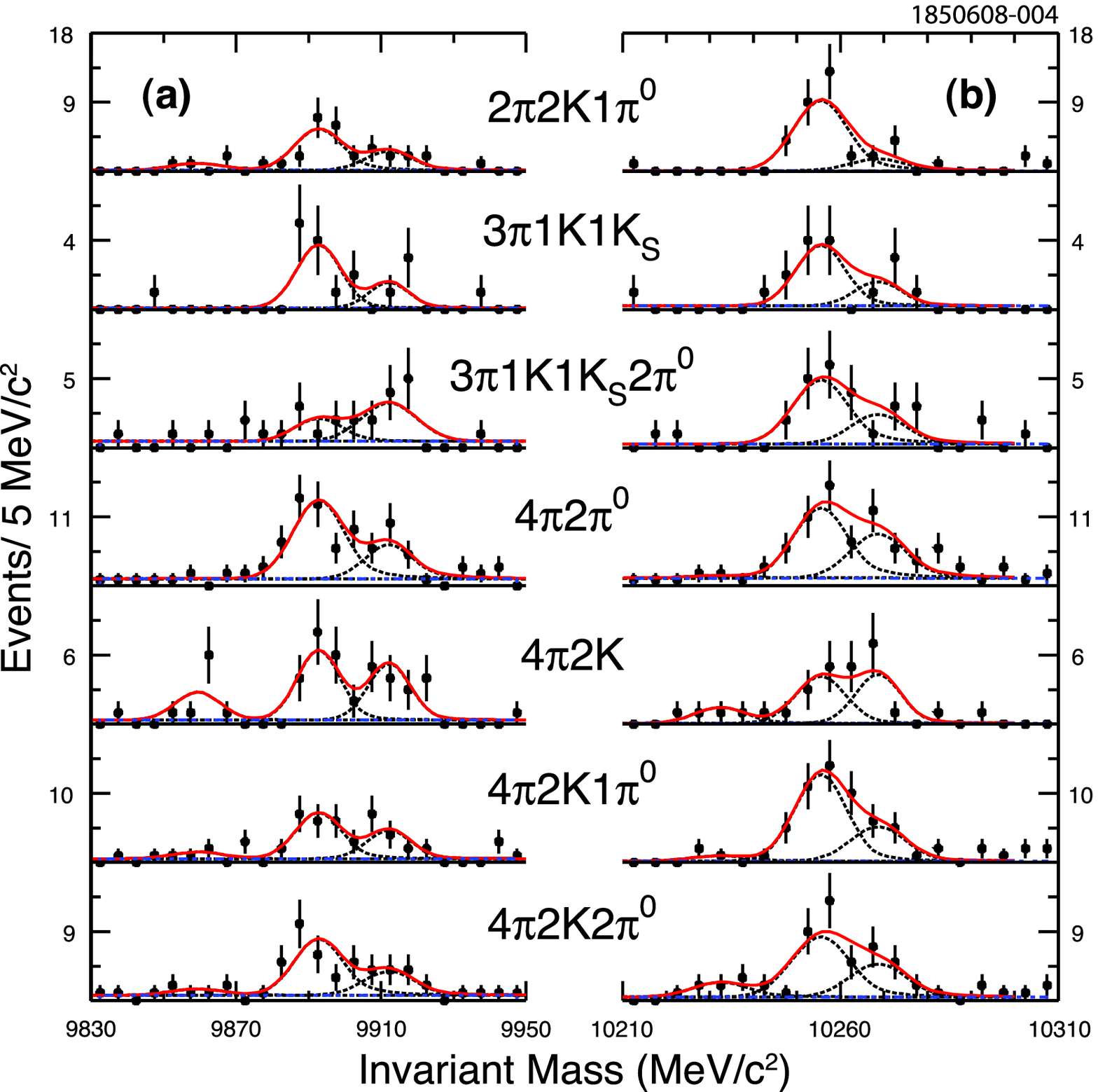}
      \includegraphics[width=0.48\textwidth]{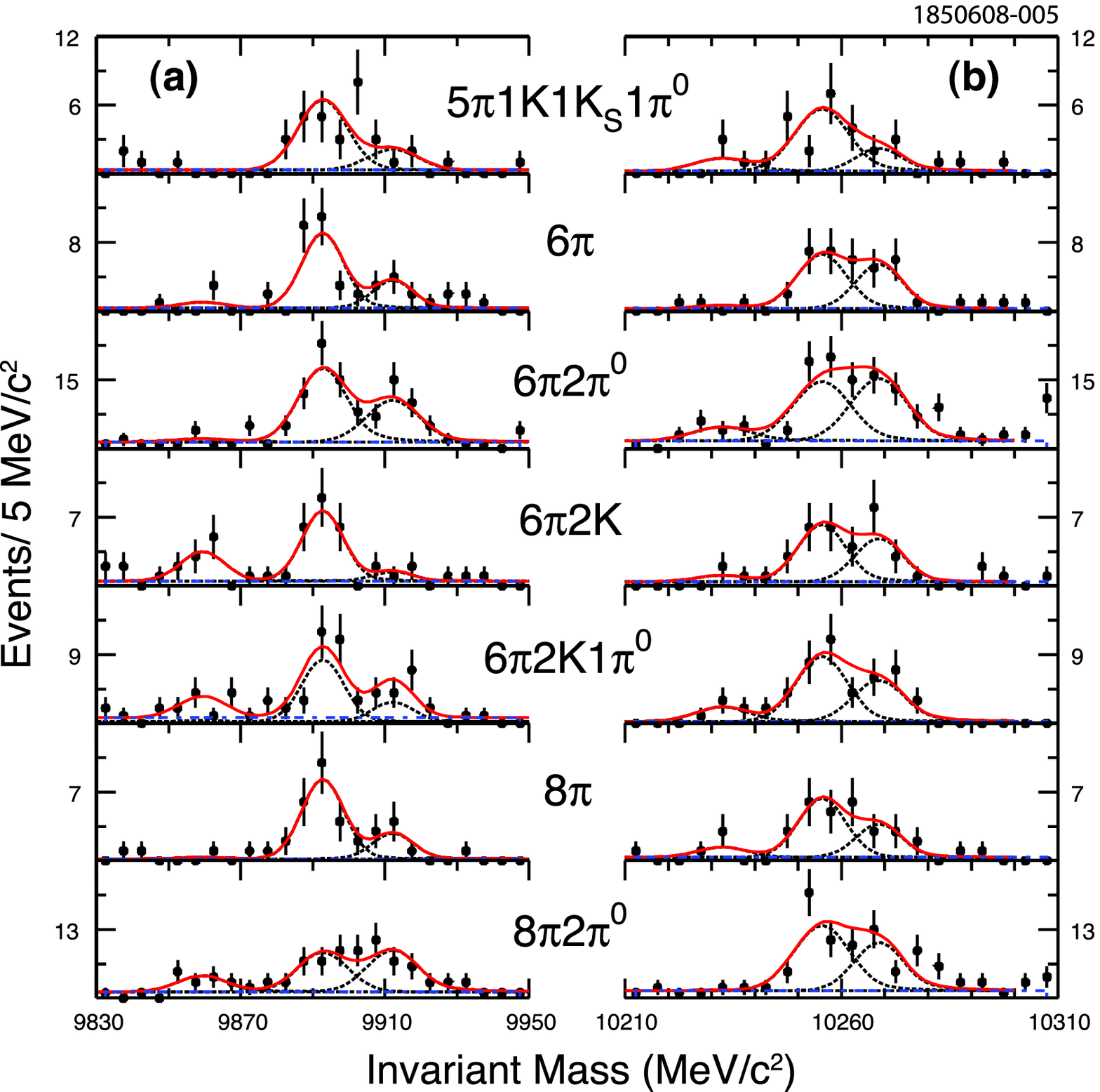}}
\caption{Fits to invariant masses of individual decay modes.
For each mode $X$, (a) refers to $\Upsilon(2S)\to\gamma\chi_b(1P)$, $\chi_b(1P)
\to X_i$, while (b) refers to $\Upsilon(3S)\to\gamma\chi_b(2P)$, $\chi_b(2P)
\to X_i$.
In labels for $X$, $K_S \equiv K_S^0$.
Dotted lines represent fitted constant backgrounds while dashed
lines are fitted $\chi_b(nP_J)$ signals of each $J$ state. 
\label{fig:chib-stack-each}}
\end{figure*}

To measure ${\cal B}[\Upsilon(nS)\to\gamma\chi_{b}((n-1)P_J)] \times
{\cal B}[\chi_{b}((n-1)P_J)\to X_i]$, where $X_i$ is each of the 14 modes,
fits to signal Monte Carlo samples for signals produced through transitions of
$\Upsilon(nS)\to\gamma\chi_b((n-1)P_J)$ are performed to $\minv$ spectra.  Once
signal shapes are fixed for each mode, we perform fits to data. We fix the
central values of $\chi_b(1P,2P)$ masses according to world averages \cite{pdg}.
Fitted $\chi^2(4C)/\dof$ distributions for each mode of $\chi_b(1P,2P)$ and
each $J$ are found to behave as expected from signal MC samples.
Resultant fits to $\minv$ spectra are shown in Fig.\
\ref{fig:chib-stack-each}.  For all cases, we use a constant (flat)
background shape and fitting ranges of 9800-9950~MeV/$c^2$ in $\Upsilon(2S)$
data and 10180-10300~MeV/$c^2$ in $\Upsilon(3S)$ data.

The major source of systematic uncertainty is found to be the effects on
signal efficiency of possible intermediate states.  We study this in
$\Upsilon(2S) \to \gamma \chi_b(1P)$ and apply the result to $\Upsilon(3S) \to
\gamma \chi_b(2P)$ as well.  In all our signal Monte Carlo samples, $\chi_b$
decays were generated according to phase space.  To estimate the systematic
uncertainty due to the presence of intermediate states, we consider
$\rho^{\pm}\to \pi^{\pm}\pi^0$, $\rho^0\to \pi^+\pi^-$, $\phi\to K^+K^-$,
$\omega\to\pi^+\pi^-\pi^0$, $\eta\to\pi^+\pi^-\pi^0$, $K^*(892)^0\to K^{\pm}
\pi^{\mp}/K^0_S\pi^0$, and $K^*(892)^{\pm}\to K^{\pm}\pi^0 /K^0_S\pi^{\pm}$.
We find deviations in the efficiencies based on these modes to be as large as
18\%; hence to allow for possible neglected intermediate states, we assign
$\pm 20\%$ systematic uncertainty due to MC modeling of $\chi_b(nP)$ decays.

Other sources of systematic errors were found to be small in comparison with
the possible presence of intermediate states.
In roughly descending order of importance, they are:
kinematic fitting (7--14\%);
photon, $\pi^0$, and charged track reconstruction (4--10\%);
particle identification and $K^0_S$ efficiencies (4--10\%);
statistical uncertainty on signal MC samples (2--8\%); 
numbers of $\Upsilon(nS)$ (2\%);
cross feeds among our 14 signal modes (1\%);
fit ranges;
background shapes;
bin width;
signal widths;
peak positions of $\chi_b$;
trigger simulation;
and
multiple candidates.
Systematic errors are added in quadrature mode by mode.  They fall within a
range of 23--30\% for all modes.

\begin{turnpage}
\begin{table}
\caption{
Reconstruction efficiencies for
$\Upsilon(nS)\to\gamma + \chi_b((n-1)P_J)\to\gamma + X$ ($\epsilon$ in units
of $10^{-2}$), event yields (N), and signal significances ($\sigma$) for each
of the transitions for each of the 14 modes.
In this and subsequent tables $\pi \equiv \pi^\pm,~K \equiv K^\pm$.
\label{tab:alleff}}
\begin{center}
\def\1#1#2#3{\multicolumn{#1}{#2}{#3}}
\begin{tabular}{l
  p{0.23in} p{0.23in} p{0.33in} p{0.23in} p{0.23in} p{0.33in} p{0.23in} p{0.23in} p{0.33in} 
  p{0.23in} p{0.23in} p{0.33in} p{0.23in} p{0.23in} p{0.33in} p{0.23in} p{0.23in}  p{0.33in}}
\hline
\hline
 & 
\1{9}{c}{$\Upsilon(2S)\to\gamma\chi_b(1P_J),\chi_b(1P_J)\to X_i$}& 
\1{9}{c}  {$\Upsilon(3S)\to\gamma\chi_b(2P_J),\chi_b(2P_J)\to X_i$} \\
\cline{2-19}
\1{1}{c}{$X_i$} & \1{3}{c}{$J=0$} & \1{3}{c}{$J=1$} & \1{3}{c}{$J=2$} & 
   \1{3}{c}{$J=0$} & \1{3}{c}{$J=1$} & \1{3}{c}{$J=2$} \\
\cline{2-19}
 & \1{1}{c}{$\epsilon$} & \1{1}{c}{ N } & \1{1}{c}{ $\sigma$ }
 & \1{1}{c}{$\epsilon$} & \1{1}{c}{ N } & \1{1}{c}{ $\sigma$ }
 & \1{1}{c}{$\epsilon$} & \1{1}{c}{ N } & \1{1}{c}{ $\sigma$ }
 & \1{1}{c}{$\epsilon$} & \1{1}{c}{ N } & \1{1}{c}{ $\sigma$ }
 & \1{1}{c}{$\epsilon$} & \1{1}{c}{ N } & \1{1}{c}{ $\sigma$ }
 & \1{1}{c}{$\epsilon$} & \1{1}{c}{ N } & \ \ $\sigma$  \\
\hline
$2\pi2K1\pi^0$     & \raggedleft$13.3$ & \raggedleft$3$  & \centering$2.0$
                   & \raggedleft$14.5$ & \raggedleft$18$ & \centering$6.4$
                   & \raggedleft$14.1$ & \raggedleft$8$  & \centering$3.4$
                   & \raggedleft$12.0$ & \raggedleft$0$  & \centering$0.0$
                   & \raggedleft$13.1$ & \raggedleft$30$ & \centering$8.9$
                   & \raggedleft$12.6$ & \raggedleft$5$  & \1{1}{c}{$2.1$}\\
$3\pi1K1K^0_S$     & \raggedleft$12.3$ & \raggedleft$0$  & \centering$0.0$
                   & \raggedleft$13.0$ & \raggedleft$11$ & \centering$5.6$
                   & \raggedleft$12.8$ & \raggedleft$4$  & \centering$2.9$
                   & \raggedleft$11.0$ & \raggedleft$0$  & \centering$0.0$
                   & \raggedleft$12.7$ & \raggedleft$10$ & \centering$4.4$
                   & \raggedleft$12.1$ & \raggedleft$4$  & \1{1}{c}{$2.1$}\\
$3\pi1K1K^0_S2\pi^0$ & \raggedleft$2.8$ & \raggedleft$1$  & \centering$0.0$
                   & \raggedleft$ 3.0$ & \raggedleft$6$  & \centering$2.2$
                   & \raggedleft$ 3.0$ & \raggedleft$11$ & \centering$3.9$
                   & \raggedleft$ 2.8$ & \raggedleft$0$  & \centering$0.0$
                   & \raggedleft$ 2.8$ & \raggedleft$15$ & \centering$5.1$
                   & \raggedleft$ 2.8$ & \raggedleft$7$  & \1{1}{c}{$2.3$} \\
$4\pi2\pi^0$       & \raggedleft$8.0$  & \raggedleft$0$  & \centering$0.1$
                   & \raggedleft$9.0$  & \raggedleft$46$ & \centering$8.5$
                   & \raggedleft$8.1$  & \raggedleft$19$ & \centering$4.3$
                   & \raggedleft$7.6$  & \raggedleft$1$  & \centering$0.5$
                   & \raggedleft$8.0$  & \raggedleft$36$ & \centering$6.7$
                   & \raggedleft$7.5$  & \raggedleft$23$ & \1{1}{c}{$4.3$} \\
$4\pi2K$           & \raggedleft$17.8$ & \raggedleft$7$  & \centering$3.5$
                   & \raggedleft$18.7$ & \raggedleft$18$ & \centering$6.3$
                   & \raggedleft$18.4$ & \raggedleft$14$ & \centering$4.9$
                   & \raggedleft$16.2$ & \raggedleft$4$  & \centering$2.4$
                   & \raggedleft$16.9$ & \raggedleft$12$ & \centering$4.6$
                   & \raggedleft$16.1$ & \raggedleft$11$ & \1{1}{c}{$4.4$} \\
$4\pi2K1\pi^0$     & \raggedleft$8.9$  & \raggedleft$3$  & \centering$1.4$
                   & \raggedleft$9.8$ & \raggedleft$22$ & \centering$6.2$
                   & \raggedleft$9.0$  & \raggedleft$13$ & \centering$4.1$
                   & \raggedleft$8.4$  & \raggedleft$2$  & \centering$1.5$
                   & \raggedleft$9.2$  & \raggedleft$38$ & \centering$8.9$
                   & \raggedleft$8.3$  & \raggedleft$16$ & \1{1}{c}{$4.2$} \\
$4\pi2K2\pi^0$     & \raggedleft$4.3$  & \raggedleft$3$  & \centering$1.1$
                   & \raggedleft$4.7$  & \raggedleft$26$ & \centering$6.2$
                   & \raggedleft$4.3$  & \raggedleft$11$ & \centering$3.2$
                   & \raggedleft$3.5$  & \raggedleft$7$  & \centering$2.5$
                   & \raggedleft$3.8$  & \raggedleft$27$ & \centering$6.5$
                   & \raggedleft$3.9$  & \raggedleft$14$ & \1{1}{c}{$3.7$} \\
$5\pi1K1K^0_S1\pi^0$ & \raggedleft$3.5$  & \raggedleft$0$  & \centering$0.0$
                   & \raggedleft$3.6$  & \raggedleft$21$ & \centering$6.3$
                   & \raggedleft$3.9$  & \raggedleft$6$  & \centering$2.4$
                   & \raggedleft$3.6$  & \raggedleft$4$  & \centering$2.2$
                   & \raggedleft$3.5$  & \raggedleft$17$ & \centering$5.5$
                   & \raggedleft$3.5$  & \raggedleft$6$  & \1{1}{c}{$2.2$} \\
$6\pi$             & \raggedleft$19.7$ & \raggedleft$2$  & \centering$1.2$
                   & \raggedleft$21.7$ & \raggedleft$25$ & \centering$7.8$
                   & \raggedleft$20.6$ & \raggedleft$9$  & \centering$3.6$
                   & \raggedleft$17.4$ & \raggedleft$1$  & \centering$0.7$
                   & \raggedleft$19.5$ & \raggedleft$18$ & \centering$5.9$
                   & \raggedleft$18.1$ & \raggedleft$14$ & \1{1}{c}{$4.7$} \\
$6\pi2\pi^0$       & \raggedleft$4.3$  & \raggedleft$3$  & \centering$0.8$
                   & \raggedleft$5.0$  & \raggedleft$56$ & \centering$9.8$
                   & \raggedleft$5.0$  & \raggedleft$34$ & \centering$6.4$
                   & \raggedleft$4.5$  & \raggedleft$11$ & \centering$2.5$
                   & \raggedleft$4.7$  & \raggedleft$44$ & \centering$7.1$
                   & \raggedleft$4.5$  & \raggedleft$45$ & \1{1}{c}{$6.7$} \\
$6\pi2K$           & \raggedleft$10.7$ & \raggedleft$9$  & \centering$3.7$
                   & \raggedleft$12.4$ & \raggedleft$21$ & \centering$6.7$
                   & \raggedleft$12.1$ & \raggedleft$3$  & \centering$1.3$
                   & \raggedleft$9.9$ & \raggedleft$2$  & \centering$1.1$
                   & \raggedleft$10.7$ & \raggedleft$16$ & \centering$5.1$
                   & \raggedleft$10.6$ & \raggedleft$12$ & \1{1}{c}{$4.2$} \\
$6\pi2K1\pi^0$     & \raggedleft$5.1$  & \raggedleft$9$  & \centering$2.9$
                   & \raggedleft$5.9$  & \raggedleft$28$ & \centering$7.3$
                   & \raggedleft$5.8$  & \raggedleft$14$ & \centering$4.3$
                   & \raggedleft$4.6$  & \raggedleft$6$  & \centering$2.9$
                   & \raggedleft$5.4$  & \raggedleft$25$ & \centering$7.0$
                   & \raggedleft$4.9$  & \raggedleft$16$ & \1{1}{c}{$5.1$} \\
$8\pi$             & \raggedleft$12.7$ & \raggedleft$0$  & \centering$0.4$
                   & \raggedleft$13.9$ & \raggedleft$24$ & \centering$7.9$
                   & \raggedleft$12.9$ & \raggedleft$7$  & \centering$3.7$
                   & \raggedleft$10.7$ & \raggedleft$3$  & \centering$1.8$
                   & \raggedleft$11.9$ & \raggedleft$16$ & \centering$5.4$
                   & \raggedleft$11.4$ & \raggedleft$9$  & \1{1}{c}{$3.2$} \\
$8\pi2\pi^0$       & \raggedleft$2.8$  & \raggedleft$11$ & \centering$2.8$
                   & \raggedleft$2.9$  & \raggedleft$26$ & \centering$5.5$
                   & \raggedleft$2.4$  & \raggedleft$29$ & \centering$5.7$
                   & \raggedleft$2.3$  & \raggedleft$0$  & \centering$0.1$
                   & \raggedleft$2.6$  & \raggedleft$41$ & \centering$7.5$
                   & \raggedleft$2.6$  & \raggedleft$27$ & \1{1}{c}{$4.8$} \\
\hline
\hline
\end{tabular}
\end{center}
\end{table}
\end{turnpage}

\begin{turnpage}
\begin{table}[htbp]
\caption{Values of ${\cal B}[\Upsilon(nS)\to\gamma\chi_b((n-1)P_J)]
\times {\cal B}[\chi_b((n-1)P_J)\to X_i]$ ($10^{-5}$).
Upper limits at
$90\%$ C.L.  are set for modes with less than $3~\sigma$ significance
(see Table~\ref{tab:alleff}).
\label{tab:BBrate}}
\begin{center}
\begin{tabular}{l  p{0.95in} p{0.95in} p{0.95in} p{0.95in} p{0.95in} p{0.95in}}
\hline
\hline
\1{1}{c}{$X_i$} & \1{2}{c}{J=0}  & \1{2}{c}{J=1} & \1{2}{c}{J=2} \\
\cline{2-7}
 & \centering$n=2$
 & \centering$n=3$
 & \centering$n=2$
 & \centering$n=3$
 & \centering$n=2$
 & \ \ \ \ \ \ $n=3$ \\
\hline
$2\pi2K1\pi^0$     & \1{1}{c}{$<0.6$} &  \1{1}{c}{$<0.2$}
                   &\1{1}{r}{$1.4\pm0.3\pm0.3$} &\1{1}{r}{$3.9\pm0.8\pm0.9$}
                   &\1{1}{r}{$0.6\pm0.3\pm0.2$} & \1{1}{c}{$<1.4$} \\
$3\pi1K1K^0_S$     & \1{1}{c}{$<0.2$} & \1{1}{c}{$<0.3$}
                   &\1{1}{r}{$0.9\pm0.3\pm0.2$}&\1{1}{r}{$1.4\pm0.5\pm0.3$}
                   & \1{1}{c}{$<0.7$} & \1{1}{c}{$<1.2$} \\
$3\pi1K1K^0_S2\pi^0$ & \1{1}{c}{$<1.8$} & \1{1}{c}{$<1.3$}  
                   & \1{1}{c}{$<4.2$} & \1{1}{r}{$9.7\pm3.0\pm2.6$}
                   &\1{1}{r}{$3.8\pm1.4\pm1.0$} & \1{1}{c}{$<8.7$} \\
$4\pi2\pi^0$       & \1{1}{c}{$<0.8$} & \1{1}{c}{$<1.4$} 
                   &\1{1}{r}{$5.5\pm0.9\pm1.4$} &\1{1}{r}{$7.4\pm1.6\pm1.9$}
                   &\1{1}{r}{$2.5\pm0.8\pm0.6$} &\1{1}{r}{$5.1\pm1.6\pm1.3$}\\
$4\pi2K$           &\1{1}{r}{$0.4\pm0.2\pm0.1$} &\1{1}{c}{$<0.9$}
                   &\1{1}{r}{$1.0\pm0.3\pm0.2$} &\1{1}{r}{$1.2\pm0.4\pm0.3$}
                   &\1{1}{r}{$0.8\pm0.2\pm0.2$} &\1{1}{r}{$1.2\pm0.4\pm0.3$}\\
$4\pi2K1\pi^0$     & \1{1}{c}{$<1.0$} & \1{1}{c}{$<1.3$}   
                   &\1{1}{r}{$2.4\pm0.6\pm0.6$} & \1{1}{r}{$6.9\pm1.3\pm1.7$}
                   &\1{1}{r}{$1.5\pm0.5\pm0.4$} &\1{1}{r}{$3.2\pm1.1\pm0.8$}\\
$4\pi2K2\pi^0$     & \1{1}{c}{$<2.0$} & \1{1}{c}{$<6.3$}  
                   &\1{1}{r}{$5.9\pm1.4\pm1.7$} &\1{1}{r}{$12.1\pm2.9\pm3.3$}
                   &\1{1}{r}{$2.8\pm1.1\pm0.7$} &\1{1}{r}{$6.2\pm2.3\pm1.7$}\\
$5\pi1K1K^0_S1\pi^0$ & \1{1}{c}{$<0.6$} & \1{1}{c}{$<3.9$}  
                   &\1{1}{r}{$6.4\pm1.6\pm1.6$} &\1{1}{r}{$8.5\pm2.3\pm2.2$}
                   & \1{1}{c}{$<3.6$} & \1{1}{c}{$<5.8$} \\
$6\pi$             & \1{1}{c}{$<0.3$} & \1{1}{c}{$<0.4$} 
                   &\1{1}{r}{$1.3\pm0.3\pm0.3$}&\1{1}{r}{$1.5\pm0.4\pm0.3$}
                   &\1{1}{r}{$0.5\pm0.2\pm0.1$}&\1{1}{r}{$1.2\pm0.4\pm0.3$}\\
$6\pi2\pi^0$       & \1{1}{c}{$<2.2$} & \1{1}{c}{$<7.2$}  
                   &\1{1}{r}{$11.9\pm1.8\pm3.2$}&\1{1}{r}{$15.0\pm3.0\pm4.0$}
                   &\1{1}{r}{$7.3\pm1.6\pm2.0$}&\1{1}{r}{$15.9\pm3.3\pm4.3$}\\
$6\pi2K$           &\1{1}{r}{$0.9\pm0.4\pm0.2$}& \1{1}{c}{$<0.9$}
                   &\1{1}{r}{$1.8\pm0.4\pm0.4$}&\1{1}{r}{$2.5\pm0.7\pm0.6$}
                   & \1{1}{c}{$<0.6$} &\1{1}{r}{$1.9\pm0.7\pm0.5$} \\
$6\pi2K1\pi^0$     & \1{1}{c}{$<3.7$} & \1{1}{c}{$<4.3$}  
                   &\1{1}{r}{$5.2\pm1.1\pm1.4$}&\1{1}{r}{$7.7\pm1.7\pm2.1$}
                   &\1{1}{r}{$2.6\pm0.8\pm0.7$}&\1{1}{r}{$5.5\pm1.6\pm1.5$} \\
$8\pi$             & \1{1}{c}{$<0.3$} & \1{1}{c}{$<1.0$} 
                   &\1{1}{r}{$1.8\pm0.4\pm0.5$}&\1{1}{r}{$2.2\pm0.6\pm0.5$}
                   &\1{1}{r}{$0.6\pm0.2\pm0.2$}&\1{1}{r}{$1.2\pm0.5\pm0.3$} \\
$8\pi2\pi^0$      & \1{1}{c}{$<7.7$}& \1{1}{c}{$<3.8$}  
                  &\1{1}{r}{$9.6\pm2.4\pm2.9$}&\1{1}{r}{$24.1\pm4.7\pm7.2$}
                  &\1{1}{r}{$13.2\pm3.1\pm4.0$}&\1{1}{r}{$16.5\pm4.6\pm5.0$}\\
\hline \hline
\end{tabular}
\end{center}
\end{table}
\end{turnpage}

\begin{turnpage}
\begin{table}[htbp]
\caption{Values of ${\cal B}[\chi_b((n-1)P_J)\to X_i]$
($10^{-4}$).
Upper limits
at $90\%$ C.L.  are set for modes with less than $3~\sigma$ significance
(see Table~\ref{tab:alleff}).
\label{tab:Brate}}
\begin{center}
\begin{tabular}{l  p{0.95in} p{0.95in} p{0.95in} p{0.95in} p{0.95in} p{0.95in}}
\hline \hline
\1{1}{c}{$X_i$} & \1{2}{c}{J=0} & \1{2}{c}{J=1} & \1{2}{c}{J=2} \\
\cline{2-7}
 & \centering$n=2$
 & \centering$n=3$
 & \centering$n=2$
 & \centering$n=3$
 & \centering$n=2$
 &  \ \ \ \ \ \ $n=3$ \\
\hline
$2\pi2K1\pi^0$     & \1{1}{c}{$<1.6$} & \1{1}{c}{$<0.3$}      
                   & \1{1}{r}{$2.0\pm0.5\pm0.5$}&\1{1}{r}{$3.0\pm0.6\pm0.8$}
                   & \1{1}{r}{$0.9\pm0.4\pm0.2$}&\1{1}{c}{$<1.1$} \\
$3\pi1K1K^0_S$     & \1{1}{c}{$<0.5$} & \1{1}{c}{$<0.5$}         
                   & \1{1}{r}{$1.3\pm0.4\pm0.3$}&\1{1}{r}{$1.1\pm0.4\pm0.3$}
                   & \1{1}{c}{$<1.2$} & \1{1}{c}{$<0.9$} \\
$3\pi1K1K^0_S2\pi^0$ & \1{1}{c}{$<4.7$} & \1{1}{c}{$<2.3$}        
                   & \1{1}{c}{$<6.1$}&\1{1}{r}{$7.7\pm2.3\pm2.2$}
                   & \1{1}{r}{$5.3\pm1.9\pm1.5$} & \1{1}{c}{$<6.7$ }\\
$4\pi2\pi^0$       & \1{1}{c}{$<2.1$} & \1{1}{c}{$<2.5$}
                   & \1{1}{r}{$7.9\pm1.4\pm2.1$}&\1{1}{r}{$5.9\pm1.2\pm1.6$}
                   & \1{1}{r}{$3.5\pm1.1\pm0.9$}&\1{1}{r}{$3.9\pm1.2\pm1.1$}\\
$4\pi2K$           & \1{1}{r}{$1.2\pm0.5\pm0.3$}& \1{1}{c}{$<1.5$}       
                   & \1{1}{r}{$1.5\pm0.4\pm0.4$}&\1{1}{r}{$0.9\pm0.3\pm0.2$}
                   & \1{1}{r}{$1.2\pm0.3\pm0.3$}&\1{1}{r}{$0.9\pm0.3\pm0.2$}\\
$4\pi2K1\pi^0$     & \1{1}{c}{$<2.7$} & \1{1}{c}{$<2.2$}        
                   & \1{1}{r}{$3.4\pm0.8\pm0.9$}&\1{1}{r}{$5.5\pm1.0\pm1.5$}
                   & \1{1}{r}{$2.1\pm0.7\pm0.5$}&\1{1}{r}{$2.4\pm0.8\pm0.7$}\\
$4\pi2K2\pi^0$     & \1{1}{c}{$<5.4$} & \1{1}{c}{$<10.8$}        
                   & \1{1}{r}{$8.6\pm2.0\pm2.4$}&\1{1}{r}{$9.6\pm2.3\pm2.8$}
                   & \1{1}{r}{$3.9\pm1.6\pm1.1$}&\1{1}{r}{$4.7\pm1.8\pm1.4$}\\
$5\pi1K1K^0_S1\pi^0$ & \1{1}{c}{$<1.7$} & \1{1}{c}{$<6.7$}         
                   & \1{1}{r}{$9.2\pm2.3\pm2.5$}&\1{1}{r}{$6.7\pm1.9\pm1.9$}
                   & \1{1}{c}{$<5.0$}& \1{1}{c}{$<4.5$} \\
$6\pi$          & \1{1}{c}{$<0.8$} & \1{1}{c}{$<0.7$}        
                & \1{1}{r}{$1.8\pm0.4\pm0.4$}&\1{1}{r}{$1.2\pm0.3\pm0.3$}
                & \1{1}{r}{$0.7\pm0.3\pm0.2$}&\1{1}{r}{$0.9\pm0.3\pm0.2$}\\
$6\pi2\pi^0$    & \1{1}{c}{$<5.9$} & \1{1}{c}{$<12.3$}         
                & \1{1}{r}{$17.2\pm2.7\pm4.8$}&\1{1}{r}{$11.9\pm2.4\pm3.4$ }
                & \1{1}{r}{$10.2\pm2.2\pm2.8$}&\1{1}{r}{$12.1\pm2.5\pm3.6$}\\
$6\pi2K$        & \1{1}{r}{$2.4\pm0.9\pm0.7$}& \1{1}{c}{$<1.5$}   
                & \1{1}{r}{$2.6\pm0.6\pm0.7$}&\1{1}{r}{$2.0\pm0.6\pm0.5$}
                & \1{1}{c}{$<0.8$}&\1{1}{r}{$1.4\pm0.5\pm0.4$} \\
$6\pi2K1\pi^0$  & \1{1}{c}{$<9.9$}& \1{1}{c}{$<7.3$}        
                & \1{1}{r}{$7.5\pm1.6\pm2.1$}&\1{1}{r}{$6.1\pm1.4\pm1.8$}
                & \1{1}{r}{$3.7\pm1.2\pm1.0$}&\1{1}{r}{$4.2\pm1.2\pm1.2$} \\
$8\pi$          & \1{1}{c}{$<0.7$} & \1{1}{c}{$<1.7$}         
                & \1{1}{r}{$2.7\pm0.6\pm0.7$}&\1{1}{r}{$1.7\pm0.5\pm0.5$}
                & \1{1}{r}{$0.8\pm0.4\pm0.2$}&\1{1}{r}{$0.9\pm0.4\pm0.3$} \\
$8\pi2\pi^0$    & \1{1}{c}{$<20.5$}& \1{1}{c}{$<6.5$}         
                & \1{1}{r}{$14.0\pm3.5\pm4.3$}&\1{1}{r}{$19.2\pm3.7\pm6.0$}
                & \1{1}{r}{$18.5\pm4.4\pm5.6$}&\1{1}{r}{$12.6\pm3.5\pm4.1$} \\
\hline
\hline
\end{tabular}
\end{center}
\end{table}
\end{turnpage}

Table~\ref{tab:alleff} shows
efficiencies, yields, and signal significances for each of the 14 modes of
$\Upsilon(2,3S)\to\gamma\chi_b(1P_J,2P_J)$.  Table~\ref{tab:BBrate} shows the
measured product branching fractions
${\cal B}[\Upsilon(nS)\to\gamma\chi_b((n-1)P_J)] \times {\cal B}
[\chi_b((n-1)P_J)\to X_i]$, for $n=2$ and 3, in units of
$10^{-5}$. For all transitions with significance less than $3 \sigma$, we set
upper limits at $90\%$ confidence level (C.L.), also shown in this table.
Table \ref{tab:Brate} shows the measured values of 
${\cal B}[\chi_b((n-1)P_J)\to
X_i]$ obtained using the values of
${\cal B}[\Upsilon(nS)\to\gamma\chi_b((n-1)P_J)]$
$=(3.8\pm0.4,6.9\pm0.4,7.15\pm0.35)\%$ for $n=2$ and
$(5.9\pm0.6,12.6\pm1.2,13.1\pm1.6)\%$ for $n=3$, and
for $J=0$, $1$, and $2$ respectively~\cite{pdg} whose
uncertainties are also included in the systematic errors.

As expected for particles of mass $\sim 10$ GeV/$c^2$, exclusive decays are
distributed over many final states. The values of
$\b[\chi_{b}((n-1)P_J) \to X_i]$
listed in Table~\ref{tab:Brate} are typically a few parts in $10^4$,
suggesting that the decay modes of these 10 GeV particles are distributed over
more than a thousand different modes, of which we have investigated 659.
Several points are worth noting.

(1) The mode with the largest branching ratio which we have identified
is $6 \pi 2 \pi^0$.  Its branching ratios from the $1P_{1,2}$ and $2P_{1,2}$
states are approximately an order of magnitude larger than those for the
$6 \pi$ mode.  Modes with charged pions and an odd number of neutral pions
are forbidden by G-parity unless subsystems contain isospin-violating decays
such as $\eta \to \pi^+ \pi^- \pi^0$.  Indeed, $6 \pi \pi^0$ and $6 \pi 3
\pi^0$ decays are not seen at a statistically significant level.  The $6 \pi 4
\pi^0$ mode involves fourteen particles, while we
consider modes with a maximum of twelve.

(2) The branching ratios for $8 \pi 2 \pi^0$ states from $1P_{1,2}$ and
$2P_{1,2}$ also exceed those for $8 \pi$ by a considerable margin.  Again,
G-parity conservation explains why one does not see a significant signal for
$8 \pi \pi^0$.

(3) Modes with one or more $K \bar K$ pairs in addition to charged pions
are exempt from the G-parity selection rule because a $K \bar K$ pair can
have either G-parity.

(4) The $4 \pi 2 \pi^0$ mode has a larger significance than either
$4 \pi$ or $4 \pi 4 \pi^0$.  Typically in the decay of an isospin-zero
particle one should expect to see the same number of $\pi^+$, $\pi^-$, and
$\pi^0$ \cite{isospin}, and this is reflected to some extent in individual
modes.

(5) The 14 modes constitute a total of less than a percent of all expected
hadronic modes of the $1P_{1,2}$ states.  The ability to identify even such
a small subset of the $1P_{1,2}$ hadronic decays depends to a large extent
on CLEO's ability to reconstruct one or more neutral pions.  Using only charged
tracks one would reconstruct an order of magnitude fewer decays.

\begin{figure}
\begin{center}
\includegraphics[width=0.48\textwidth]{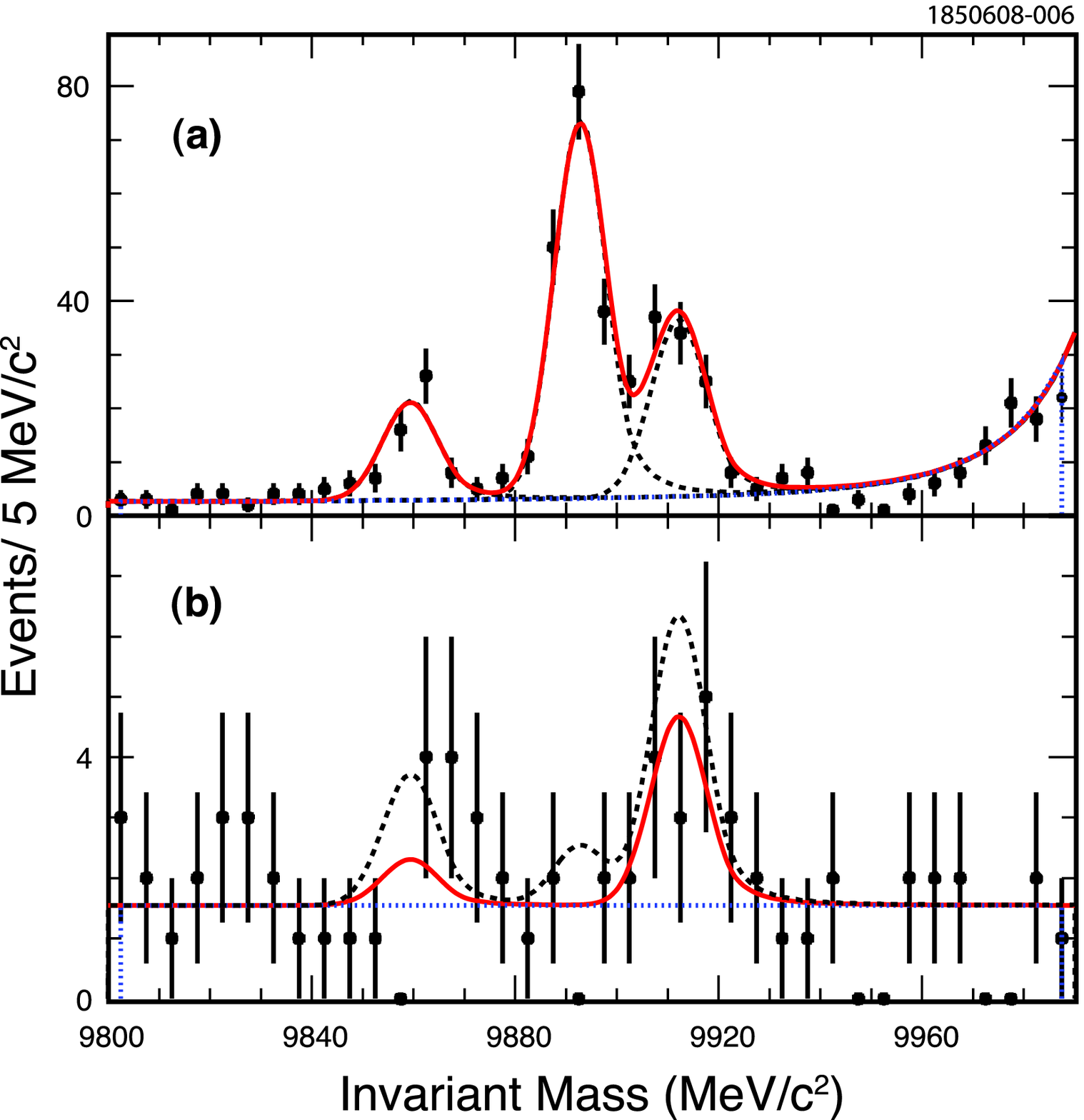}
\caption{Comparison of spectra, based on 167 modes not involving $\pi^0$ or
$\eta$ mesons, in transitions from (a) $\Upsilon(2S) \to \gamma \chi_b(1P)$
and (b) $\Upsilon(3S) \to \gamma \chi_b(1P)$.  The background shape in (a) is
derived from $\Upsilon(1S)$ data as described in the text.
The dashed lines in (a) are fitted $\chi_b(nP_J)$ signals of each $J$ state
while those in (b) correspond to $90\%$ C.L. upper limits on signal yields.}
\label{fig:suppe1}
\end{center}
\end{figure}

We have also studied the E1 transitions $\Upsilon(3S) \to \gamma \chi_b(1P_J)$.
These transitions are suppressed by small overlaps of wave functions in the
dipole matrix element $\langle 1P | \vec{r} | 3S \rangle$ \cite{E1supp}.  We
investigate the ratios
\beq \label{eqn:RJ}
R_J \equiv \frac{\b[\Upsilon(3S)\to\gamma\chi_b(1P_J)]}
                {\b[\Upsilon(2S)\to\gamma\chi_b(1P_J)]}~,
\eeq
where the ratios of branching fractions are determined from fitted yields in
$\Upsilon(3S)$ and $\Upsilon(2S)$ data, respectively, corrected for small
differences in signal efficiencies.  In modes with any neutral pions or $\eta$
mesons, substantial backgrounds arise from subsequent 
$\chi_b(1P_J) \to \gamma \Upsilon(1S)$ decays,
whose photons are similar in energy to those in
$\Upsilon(3S) \to \gamma \chi_b(1P_J)$.  To eliminate such backgrounds, we
restrict attention to 167 modes not involving $\pi^0$ or $\eta$ mesons.
Fig.\ \ref{fig:suppe1}(a) shows the $\minv$ distribution based on the sum of
such modes in $\Upsilon(2S)$ data.  The background is represented by the
exponential of a polynomial, fitted to a {\it shifted} invariant mass
distribution of $\Upsilon(1S)$ data (smooth dotted curve), and the peak
positions are fixed at the known masses~\cite{pdg}.

Using the $\chi_b(1P_J)$ signal shapes obtained in this fit, we fit
$\Upsilon(3S)$ data with a constant background, as shown in
Fig.\ \ref{fig:suppe1}(b).  The fit gives statistical significances of $0.7
\sigma$, $0.0 \sigma$, and $2.6 \sigma$ for signals consistent with the
transitions $\Upsilon(3S) \to \gamma \chi_b(1P_{0,1,2})$, respectively. We find
$R_2 = (15.7 \pm 7.6 \pm 2.2) \times 10^{-2}$ and 90\% C.L.\ upper limits
$R_{(0,1,2)} < (21.9,2.5,27.1) \times 10^{-2}$.  Using known values of
$\b[\Upsilon(2S) \to \gamma \chi_b(1P)]$ \cite{pdg}, we then find
$\b[\Upsilon(3S) \to \gamma \chi_b(1P_2)] = [11 \pm 6({\rm stat.}) \pm 2({\rm
syst.}) \pm 1] \times 10^{-3}$, where the last uncertainty comes from
$\b[\Upsilon(2S)\to\gamma\chi_b(1P_2)]$~\cite{pdg}.  While most systematic
uncertainties are canceled in the ratio of yields, our total uncertainty
($14\%$) is dominated by variations in the signal as we change the width of the
range over which we fit $\Upsilon(3S)$ decays.
Our nominal fit range is 9800-9990~MeV, varied to 9800-9950~MeV
and 9750--10050~MeV.  Although this variation is well within statistical
fluctuations, we conservatively take it as a possible systematic uncertainty.

We set 90\% C.L.\ upper limits $\b[\Upsilon(3S) \to \gamma \chi_b(1P_0)] <
9.2 \times 10^{-3}$, consistent with the value of $(3.0 \pm 0.4 \pm 1.0) 
\times 10^{-3}$ reported in Ref.\ \cite{Artuso:2004fp}, $\b[\Upsilon(3S) \to
\gamma \chi_b(1P_1)] < 1.9 \times 10^{-3}$, and $\b[\Upsilon(3S) \to \gamma
\chi_b(1P_2)] < 20.3 \times 10^{-3}$. 
Our results are compared with
some theoretical predictions in Table \ref{tab:hinderedrateth}.

We have presented the first observations of decays of $\chi_b(1P_J)$
and $\chi_b(2P_J)$ to exclusive final states of light hadrons.  These results
can be of use in validating models for fragmentation of heavy states, and in
searching for states of mass $~\sim 10$ GeV/$c^2$ via their exclusive decays.
We also find upper limits for the rates of the suppressed E1
transitions $\Upsilon(3S) \to \gamma \chi_b(1P_{0,1,2})$.

We gratefully acknowledge the effort of the CESR staff
in providing us with excellent luminosity and running conditions.
D.~Cronin-Hennessy and A.~Ryd thank the A.P.~Sloan Foundation.
This work was supported by the National Science Foundation,
the U.S. Department of Energy,
the Natural Sciences and Engineering Research Council of Canada, and
the U.K. Science and Technology Facilities Council.

\begin{table}
\begin{center}
\caption{
Comparison of measurements and predictions \cite{hinde1th} for
suppressed E1 transition rates in units of eV.  Experimental
measurements
are based on $\Gamma_{\Upsilon(3S)} = 20.32$~keV~\cite{pdg}.
\label{tab:hinderedrateth}}
\begin{tabular}{ l   p{0.5in}   p{0.5in}   p{0.5in} }
\hline
\hline
 & \centering$J=0$ & \centering$J=1$ & \ \ $J=2$ \\
\hline
Inclusive expt.\ ~\cite{Artuso:2004fp} & \1{1}{c}{$61\pm22$} & \1{1}{c}{-} &
 \1{1}{c}{-} \\
Exclusive expt.\ (this work) & \1{1}{c}{$<186$} & \1{1}{c}{$<38$} &
 \1{1}{c}{$<413$} \\
Moxhay--Rosner (1983) & \1{1}{c}{$25$} & \1{1}{c}{$25$} & \1{1}{c}{$150$} \\
Grotch {\it et al.} (1984) (a) & \1{1}{c}{$114$} & \1{1}{c}{$3.4$} & 
  \1{1}{c}{$194$} \\
Grotch {\it et al.} (1984) (b) & \1{1}{c}{$130$} & \1{1}{c}{$0.3$} &
  \1{1}{c}{$430$} \\
Daghighian--Silverman (1987) & \1{1}{c}{$16$} & \1{1}{c}{$100$} & 
  \1{1}{c}{$650$} \\
Fulcher (1990) & \1{1}{c}{$10$} & \1{1}{c}{$20$} & \1{1}{c}{$30$} \\
L\"ahde (2003) & \1{1}{c}{$150$} & \1{1}{c}{$110$} & \1{1}{c}{$40$} \\
Ebert {\it et al.} (2003) & \1{1}{c}{$27$} & \1{1}{c}{$67$} & \1{1}{c}{$97$} \\
\hline
\hline
\1{4}{l}{(a) Confining potential is purely scalar.} \\
\1{4}{l}{(b) Confining potential is purely vector.} \\
\end{tabular}
\end{center}
\end{table}

\end{document}